\pdfoutput=1

\documentclass[11pt]{article}

\usepackage{acl}

\usepackage{microtype}
\usepackage{hyperref}
\usepackage{url}
\usepackage{booktabs}

\usepackage{times}
\usepackage{latexsym}

\usepackage[T1]{fontenc}
\usepackage[utf8]{inputenc}

\usepackage{enumerate}
\usepackage{graphicx}
\usepackage{subcaption}
\usepackage{graphics, amsmath}
\usepackage{multirow}

\usepackage{tabularx}
\usepackage{pifont}
\usepackage{colortbl}

\usepackage{xcolor}
\usepackage[inline]{enumitem}
\usepackage{makecell}
\usepackage{adjustbox}

\usepackage{acronym}

\usepackage{siunitx}
\usepackage{array}
\usepackage{caption}
\usepackage{wrapfig}
\usepackage{xspace}

\usepackage{soul}
\usepackage{arydshln}
\usepackage{stfloats}

\definecolor{OliveGreen}{rgb}{0,0.6,0}
\definecolor{myred}{HTML}{F8CCCC}
\definecolor{mygreen}{HTML}{E0ECD4}

\newcommand{\gense}{GenSearch\xspace}

\newcommand{\factcc}{FactCC\xspace}
\newcommand{\bertscore}{BERTScore\xspace}
\newcommand{\bartscore}{BARTScore\xspace}
\newcommand{\summac}{SummaC\xspace} %
\newcommand{\summacconvfull}{SummaC\textsubscript{Conv}\xspace}
\newcommand{\summaczsfull}{SummaC\textsubscript{ZS}\xspace}

\newcommand{\autoais}{AutoAIS\xspace}
\newcommand{\alignscore}{AlignScore\xspace}

\newcommand{\gptdis}{GPT-3.5-DIS\xspace}
\newcommand{\gptcon}{GPT-3.5-CON\xspace}

\newcommand{\isfullno}{FS-vs-NS\xspace}
\newcommand{\ispartialfull}{FS-vs-PS\xspace}
\newcommand{\ispartialno}{PS-vs-NS\xspace}

\title{Towards Fine-Grained Citation Evaluation in Generated Text: \\ A Comparative Analysis of Faithfulness Metrics}

\author{Weijia Zhang$^1$ \quad Mohammad Aliannejadi$^1$ \quad Yifei Yuan$^2$ \quad Jiahuan Pei$^3$  \\
{\bf Jia-Hong Huang$^1$ \quad Evangelos Kanoulas$^1$ } \\
$^1$University of Amsterdam \quad
$^2$University of Copenhagen \\
$^3$Centrum Wiskunde \& Informatica \\
\texttt{w.zhang2@uva.nl}}

\begin{document}
\maketitle

\begin{abstract}

Large language models (LLMs) often produce unsupported or unverifiable content, known as ``hallucinations.'' To mitigate this, retrieval-augmented LLMs incorporate citations, grounding the content in verifiable sources. Despite such developments, manually assessing how well a citation supports the associated statement remains a major challenge. Previous studies use faithfulness metrics to estimate citation support automatically but are limited to binary classification, overlooking fine-grained citation support in practical scenarios. To investigate the effectiveness of faithfulness metrics in fine-grained scenarios, we propose a comparative evaluation framework that assesses the metric effectiveness in distinguishing citations between three-category support levels: \textit{full}, \textit{partial}, and \textit{no} support.
Our framework employs correlation analysis, classification evaluation, and retrieval evaluation to measure the alignment between metric scores and human judgments comprehensively. Our results show no single metric consistently excels across all evaluations, revealing the complexity of assessing fine-grained support. Based on the findings, we provide practical recommendations for developing more effective metrics.

\end{abstract}

\begin{figure}[tb]
    \centering
    \includegraphics[width=0.985\linewidth]{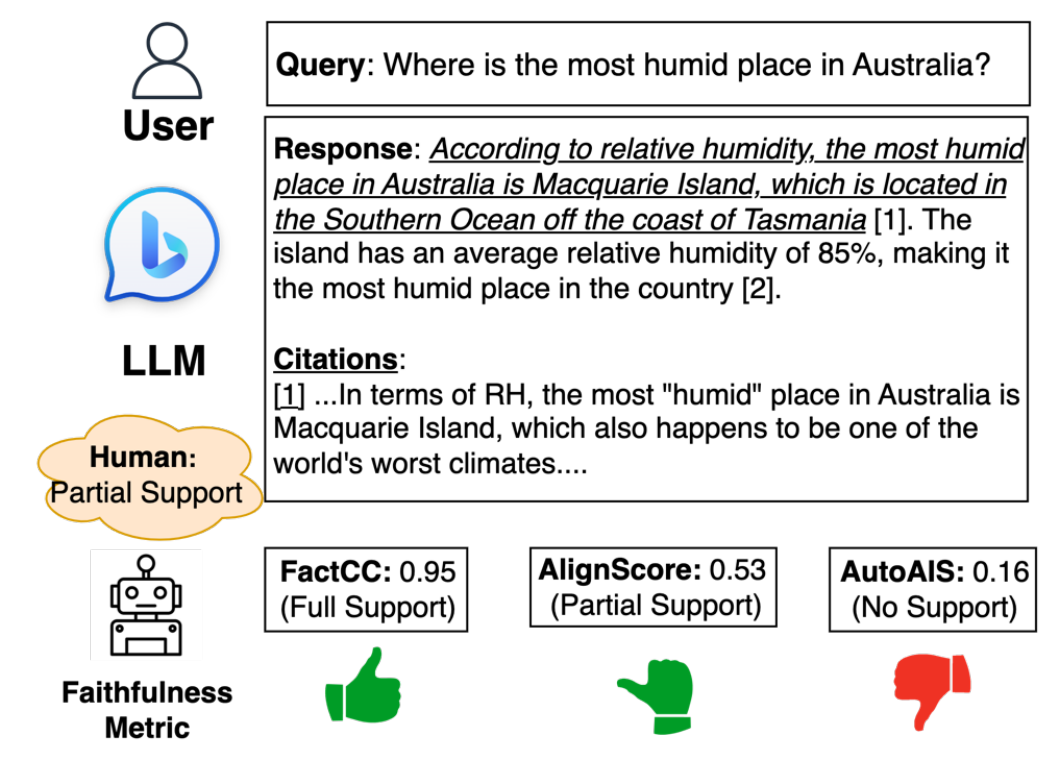}
    \caption{An example of \textit{partial support} in citation evaluation. Inconsistent metric scores are observed when assessing the statement with three faithfulness metrics.
    }
    \label{fig:example}
\end{figure}

\section{Introduction}

Large language models (LLMs) often generate content known as ``hallucinations''~\cite{li2022faithfulness,ji2022hallucination,zhang2023hallucination}, which contradicts established knowledge or lacks verification from reliable sources. Mainstream studies~\cite{bohnet2022attributed,gao-etal-2023-rarr} aim to mitigate this by using retrieval-augmented LLMs to generate responses with in-line citations that provide supporting evidence. One primary challenge is to assess how well a citation supports its statement, as manual evaluation is labor-intensive and time-consuming. Automated citation evaluation has been explored to reduce reliance on human assessments~\cite{gao-etal-2023-enabling,li2023llatrieval}. To this end, faithfulness evaluation metrics are employed as proxies to automatically estimate the citation support~\cite{xia2024ground,li-etal-2024-improving-attributed}. These metrics measure the faithfulness between model-generated and sourced text, which aligns closely with the objectives of automated citation evaluation.

Prior studies in faithfulness metrics have primarily limited this task to a binary classification problem~\cite{tahaei-etal-2024-efficient,huang-etal-2024-learning}, where faithfulness metrics are leveraged to determine whether a citation supports the associated statement. 
However, this binary approach fails to capture the fine-grained citation support encountered in real-world applications. 
For instance, in Figure \ref{fig:example}, a retrieval-augmented LLM generates a response with multiple citations given a query. 
A human assessor labels the first citation as  \textit{``partial support''} since it only supports ``the most humid place in Australia is Macquarie Island'' but not ``which is located in the Southern Ocean off the coast of Tasmania.'' This partial support scenario causes noticeable inconsistencies across three different faithfulness metrics. Therefore, there is a significant research need to evaluate the effectiveness of faithfulness metrics in accurately distinguishing citations in such fine-grained support scenarios.

To address this issue, we propose a comparative evaluation framework for assessing the metric effectiveness in fine-grained support scenarios. 
In our framework, we define ``\textit{support levels}'' as the extent to which a citation supports the associated statement~\cite{liu-etal-2023-evaluating,yue-etal-2023-automatic}. 
Specifically, we consider a three-category support level scenario: \textit{full}, \textit{partial}, and \textit{no} support. These categories indicate whether a citation fully, partially or does not support the associated statement, respectively.
To comprehensively assess the metric effectiveness, we measure the alignment between metric scores and human judgments with three types of evaluation protocols:
\begin{enumerate*}[label=\arabic*)]
    \item \textit{Correlation analysis:} we employ it to measure how well metric scores correlate with human judgments. 
    \item \textit{Classification evaluation:} we conduct a classification evaluation to assess the metrics' capability to distinguish citations based on their support levels. 
    \item \textit{Retrieval evaluation:} we undertake a retrieval evaluation to assess the metric effectiveness in ranking citations according to their support levels. This is motivated by the observation that the previous two evaluation protocols assume citations are within statements, which is not always valid in practice~\cite{asai2024selfrag}. In such cases, faithfulness metrics are adapted to perform post-hoc retrieval, aiming to retrieve potential citations from a candidate pool~\cite{kang2023ever,gou2024critic}. Thus, retrieval evaluation is crucial for determining the practical utility of these metric adaptations.
\end{enumerate*}

In our experiments, we assess various widely used faithfulness metrics, categorizing them into \textit{similarity-based}, \textit{entailment-based}, and \textit{LLM-based} metrics. We find that:  
\begin{enumerate*}[label=\arabic*)]
    \item No single faithfulness metric consistently outperforms others across three evaluation protocols, suggesting that these protocols are complementary and should be integrated for a comprehensive evaluation of metric performance;
    \item The best-performing metrics show promise in distinguishing some support scenarios but struggle with others. This highlights the inherent complexities of automated citation evaluation.
    \item 
    Similarity-based metrics surpass best-performing entailment-based metrics in retrieval evaluation. This indicates that entailment-based metrics exhibit higher sensitivity to noisy data, which is introduced by irrelevant documents in such scenarios.
\end{enumerate*}

\vspace{+3pt}
\noindent 
Our contributions can be summarized as follows:
\begin{itemize}[noitemsep,leftmargin=10px,nosep]
    \item 
    To the best of our knowledge, we are the first to systematically investigate the effect of fine-grained support levels on faithfulness metrics in the task of automated citation evaluation. 
    \item  
    We propose a comparative evaluation framework to assess the alignment between metric scores and human judgments. This framework includes three evaluation protocols to comprehensively evaluate the metric performance.
    \item  
    Our experimental results demonstrate the best-performing faithfulness metrics still struggle to identify partially supporting citations, underscoring the inherent challenges of automated citation evaluation. Based on our findings, we offer practical recommendations for the development of more effective metrics.
\end{itemize}

\begin{figure*}[t]
    \centering
    \includegraphics[width=0.985\linewidth]{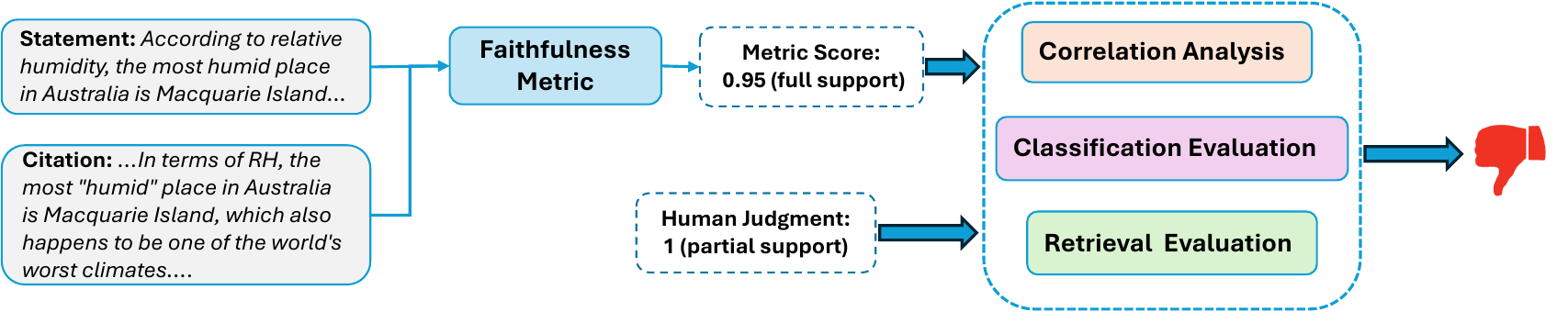}
    \caption{The overview of the proposed comparative evaluation framework. A faithfulness metric assigns scores to given statements and their corresponding citations. Subsequently, our framework comprehensively assesses the alignment between these metric scores and human judgments by employing correlation analysis, classification, and retrieval evaluation.}
    \label{fig:framework}
\end{figure*}

\section{Related Work}

\paragraph{Faithfulness Evaluation Metrics}

Faithfulness evaluation metrics are crucial for assessing the factual consistency of text generated by models relative to the source text.
It receives great interest within the field of natural language generation (NLG)~\cite{huang2019novel,huang2021deepopht,zhang2021scaling,huang2022causal,huang2023causalainer,zhang2023tackling,huang2024optimizing,huang2024novel,zhu2024enhancing}, particularly in abstractive summarization~\cite{maynez-etal-2020-faithfulness,kryscinski-etal-2020-evaluating,huang2020query,huang2021gpt2mvs,zhang2024beyond,zhang2024qfmts}.
In general, faithfulness metrics are categorized into three types: entailment-based, similarity-based, and QA-based metrics. 
Entailment-based metrics employ natural language inference (NLI) models to determine if the source text entails the generated text~\cite{falke-etal-2019-ranking,laban-etal-2022-summac,honovich-etal-2022-true-evaluating,zha-etal-2023-alignscore}. 
Similarity-based metrics, such as BERTScore~\cite{zhang2020bertscore} and BARTScore~\cite{yuan2021bartscore}, quantify text similarity and have demonstrated robust performance in faithfulness evaluation~\cite{pagnoni-etal-2021-understanding,honovich-etal-2022-true-evaluating}.
QA-based metrics utilize a combination of question generation and question answering to estimate faithfulness levels~\cite{durmus-etal-2020-feqa,wang-etal-2020-asking,scialom-etal-2021-questeval,fabbri-etal-2022-qafacteval}.
In this work, we exclude QA-based metrics from our work, following recent works suggesting the challenging limitations in these metrics~\cite{kamoi-etal-2023-shortcomings}. %
We focus on the extrinsic evaluation of faithfulness metrics against human judgments in scenarios requiring fine-grained citation support.

\paragraph{Citation Evaluation}

Citation evaluation seeks to enhance the trustworthiness of retrieval-augmented LLMs by verifying the support provided by citations to the generated statements~\cite{rashkin2021attribution,yue-etal-2023-automatic,huang2023citation,huang2024traincitation,zhang2024comparative}.
Given the labor-intensive nature of manual citation evaluation, there has been a shift towards automated approaches to reduce dependence on human evaluation.
Since the goals of automated citation evaluation align closely with faithfulness evaluation in NLG, faithfulness metrics are employed to verify whether a citation supports the corresponding statement~\cite{li2023towards,sun2023towards,ye-etal-2024-effective,li2024attrbench,shen2024citekit,huang-etal-2024-learning}. 
Despite their widespread usage, the effectiveness of these metrics in more practical fine-grained citation support scenarios, such as those involving partial support by citations, has not been adequately addressed. Questions remain about the metrics' capability to differentiate citations in these fine-grained scenarios.
This work addresses these gaps by examining the effectiveness of faithfulness metrics across three distinct levels of citation support: full, partial, and no support.

\section{Evaluation Framework}

In this section, we introduce the proposed comparative evaluation framework. We begin by formalizing the task of automated citation evaluation. Subsequently, we detail three distinct evaluation protocols within this framework, ensuring a comprehensive assessment in alignment between faithfulness metrics and human judgments. Our framework is demonstrated in Figure \ref{fig:framework}.

\subsection{Task Formulation} 

The objective of automated citation evaluation is to automatically quantify the support level of a citation based on the citation and its associated statement.
In this work, we assume access to a dataset for automated citation evaluation, comprising pairs of statements and their corresponding citations, denoted as $(s_i,c_i)$. Each $s_i$ is a statement from the set $S$ of all statements produced by an LLM and each $c_i$ is a citation from a set $C$ of citations returned by the LLM.
We categorize the citations into three distinct support levels: full, partial, and no support. We adopt the definition of these support levels from \citet{liu-etal-2023-evaluating}:
\begin{itemize}[noitemsep,leftmargin=10px,nosep]
    \item Full Support (\textsc{FS}): the citation fully supports every detail in the statement.
    \item Partial Support (\textsc{PS}): the citation supports certain aspects of the statement, while other details remain unsupported or are contradicted.
    \item No Support (\textsc{NS}): none of the content in the statement is supported by the citation. For instance, the citation is entirely irrelevant or contradicts the statement.
\end{itemize}
\noindent
To this end, without loss of generality, we define a faithfulness metric as a scoring function, denoted as $F(s_i,c_i) \rightarrow R^+$. 
For any given statement $s_i$ and its associated citation $c_i$, this scoring function provides a numeric score that indicates the extent of support provided by the citation to the statement.

\subsection{Evaluation Protocols}

The objective of evaluation protocols is to comprehensively assess the extent to which metric scores align with human judgments. In this work, we assess this alignment across three distinct dimensions: \textbf{correlation}, \textbf{classification performance}, and \textbf{retrieval effectiveness}.

\subsubsection{Correlation Analysis} 

The correlation analysis measures the general trend in the relationship between metric scores and human judgments.
Previous research~\cite{kryscinski-etal-2020-evaluating,pagnoni-etal-2021-understanding} has employed correlation analysis to meta-evaluate faithfulness metrics in abstractive text summarization. 
They involve measuring the extent to which metric scores align with binary levels of faithfulness, which are annotated by human assessors as either faithful ($1$) or unfaithful ($0$). 
Inspired by them, we adapt correlation analysis to the task of automated citation evaluation. 
Specifically, given the statements and their associated citations, we assess how well predicted metric scores correlate with human-annotated support levels. 
To facilitate correlation analysis, we  assign support levels $\{\textsc{FS}, \textsc{PS}, \textsc{NS}\}$ to values $\{0, 1, 2\}$. We then utilize standard correlation metrics to assess metric performance. The details are shown in Section \ref{subsec:meta_eval}.

\subsubsection{Classification Evaluation} 

In addition to correlation analysis, we perform classification evaluation to determine the metric effectiveness in discriminating citations based on their support level. 
Specifically, the metrics need to categorize a citation into one of three support levels: \textsc{FS}, \textsc{PS}, \textsc{NS}. 
Notably, existing faithfulness metrics do not apply to this three-way classification scenario, as they are unable to accurately determine the extent to which a statement is partially supported by its corresponding citation~\cite{laban-etal-2022-summac}. 
To address this issue, we adopt a one-vs-one strategy, by effectively decomposing the three-way classification into three binary classification task settings:
\begin{enumerate*}[label=(\roman*)]
    \item Full Support vs.\ No Support (\isfullno{}),
    \item Full Support vs.\ Partial Support (\ispartialfull{}), and
    \item Partial Support vs.\ No Support (\ispartialno{}).
\end{enumerate*}
For each binary classification task setting, we construct a specialized dataset comprising only instances with the corresponding binary support levels derived from the original dataset. 
We assess the performance of metrics on these tailored binary datasets using standard binary classification evaluation metrics. 
The overall metric performance is then computed by averaging the results across all binary tasks.

\subsubsection{Retrieval Evaluation} 

The objective of retrieval evaluation is to measure the metric effectiveness in ranking citations according to their support levels.
This evaluation is motivated by the observation that previous correlation and classification evaluations presuppose the presence of citations within generated statements. 
However, real-world scenarios frequently present instances where citations are absent or irrelevant, highlighting the need for post-hoc retrieval to enhance citation quality~\cite{liu-etal-2023-evaluating,huang2024traincitation}.
In post-hoc retrieval, candidate documents are retrieved to form a pool of potential citations using information retrieval techniques~\cite{karpukhin-etal-2020-dense}. 
Faithfulness metrics are then employed to rank citations based on their predicted metric scores, aiming to identify the citation with the highest support level.
Ideally, a faithfulness metric should rank fully supporting citations at the top, followed by partially supporting citations, and finally non-supporting citations.
Similar to correlation analysis, we assign support levels $\{\textsc{FS}, \textsc{PS}, \textsc{NS}\}$ to relevance labels $\{2, 1, 0\}$. 
The metric effectiveness is assessed using standard information retrieval evaluation metrics. This evaluation also provides a deeper understanding of metric performance in post-hoc citation retrieval scenarios.
\section{Faithfulness Metrics} 

In our experiments, we evaluate diverse faithfulness evaluation metrics, dividing them into similarity-based, entailment-based, and LLM-based metrics. Similarity-based metrics assess the support levels mainly based on the degree of similarity between the citation and the associated statement. Entailment-based metrics leverage pre-trained NLI models to estimate the support levels. 
LLM-based metrics directly prompt LLMs to measure the support levels.

\subsection{Similarity-Based Metrics}

\smallskip\noindent%
\textbf{\bertscore{}}
\cite{zhang2020bertscore} adopts BERT~\cite{devlin2019bert} to measure semantic similarity between a pair of text by aggregating cosine similarity among token-level BERT representation without further fine-tuning. 
We report the precision version of \bertscore since it correlates more with human judgments in faithfulness evaluation~\cite{pagnoni-etal-2021-understanding},  We use recommended \texttt{deberta-xlarge-mnli} \cite{he2021deberta} as the backbone model.

\smallskip\noindent%
\textbf{\bartscore{}}
\cite{yuan2021bartscore} adopts BART~\cite{lewis2020bart} to measure the similarity between two texts based on conditional log-likelihood of generating target text from source text.
In our experiments, we leverage the faithfulness version of \bartscore, in which we treat the citation and the statement as the source and target text, respectively. We use the BART model fine-tuned on the CNN/DailyMail dataset~\cite{hermann2015cnndm} as the backbone model.

\subsection{Entailment-Based Metrics}

\smallskip\noindent%
\textbf{\factcc{}} 
\cite{kryscinski-etal-2020-evaluating} is a BERT-based model to verify whether a generated text is faithful to a source text, which is fine-tuned on synthetic training data containing simulated examples with different factual errors~\cite{kryscinski-etal-2020-evaluating}.

\smallskip\noindent%
\textbf{\summac{}} 
~\cite{laban-etal-2022-summac} is a RoBERTa-based model \cite{liu2019roberta} fine-tuned on NLI datasets. This metric splits source and generated texts into sentences, computes entailment scores for each pair, and aggregates these scores to obtain the final faithfulness score.
It has two variants: 
\begin{enumerate*}[label=(\roman*)]
    \item \summaczsfull{} is a zero-shot version that is only pre-trained on NLI datasets;
    \item \summacconvfull{} adds extra convolutional layers and is further fine-tuned on synthetic training data proposed in~\citet{kryscinski-etal-2020-evaluating}. 
\end{enumerate*}

\smallskip\noindent%
\textbf{\autoais{}}~\cite{honovich-etal-2022-true-evaluating}
is a T5-11B~\cite{raffel2020exploring} model trained on a collection of NLI datasets, which is commonly used in recent automated citation evaluation. As the original output of \autoais{} is a numeric, either ``1'' (faithful) or ``0'' (unfaithful), we utilize the generated token probability of ``1'' as the predicted metric score.

\smallskip\noindent%
\textbf{\alignscore{}}~\cite{zha-etal-2023-alignscore} further fine-tunes a RoBERTa-based model~\cite{liu2019roberta} with a unified alignment loss function. To this end, a unified dataset containing a variety of related natural language processing datasets has been collected. In this work, we adapt the \texttt{large} version as it demonstrates the best performance. %

\subsection{LLM-Based Metrics}

In addition to established faithfulness metrics, we utilize LLMs as faithfulness evaluators for comparison. Specifically, we introduce two prompting methods as follows:
\begin{enumerate*}[label=(\roman*)]
    \item \textbf{Discrete scoring} prompts the LLM to assign discrete scores from the set ${0, 1, 2}$ for a given statement and its citation, where $0$, $1$, and $2$ indicate no support, partial support, and full support, respectively;
    \item \textbf{Continuous scoring} prompts the LLM to assign continuous scores in the range $[0, 1]$ for a given statement and its citation. Here, $1$ indicates full support, $0$ indicates no support, and values between 0 and 1 indicate partial support.
\end{enumerate*}

In the experiments, we employ the latest version of GPT-3.5 (\texttt{gpt-3.5-turbo-0125}) as the base model. Moreover, we utilize the chain of thought (CoT) method~\cite{wei2022cot,kojima2022zerocot} to enhance the reasoning capabilities of the LLM. We use \gptdis{} and \gptcon{} to denote GPT-3.5 using discrete and continuous scoring methods, respectively. The detailed prompts are shown in Appendix \ref{app_sec:appendix_prompts}.

\begin{table}[t]
\centering
\footnotesize
\setlength\tabcolsep{5pt}
\centering
\small
\begin{tabular}{@{}lc@{}}
\toprule[1pt]
\textbf{Human Judgment} & \textbf{\# Statement-Citation Pair}  \\
\midrule
Full Support        & 6,616         \\
Partial Support      & 1,445           \\
No Support         & 4,620           \\
\midrule
Total         & 12,681     \\
\bottomrule[1pt]
\end{tabular}
\caption{Data statistics of the \gense{} dataset. Each pair has been annotated by human assessors based on three categories: full, partial, and no support.}
\label{tab:data_stats}
\end{table}

\section{Experiments}
\label{sec:experiments}

In this section, we describe the dataset used in the experiments.
Subsequently, we discuss the evaluation metrics incorporated within our proposed framework, which assess the performance of faithfulness metrics in alignment with human judgments.

\subsection{Datasets}
\label{subsec:datasets}

In our experiments, we utilize the \gense{} dataset~\cite{liu-etal-2023-evaluating} as our evaluation benchmark, which consists of data from generative search engines (GSE) like BingChat.\footnote{\url{https://www.bing.com/chat}} These GSEs represent commercial applications of retrieval-augmented LLMs.
As depicted in Figure~\ref{fig:example}, each example includes a user query and a corresponding response generated by the GSE. The user queries are sourced from various QA datasets~\cite{fan2019eli5,kwiatkowski-etal-2019-natural}. Each response consists of multiple statements, each containing in-line citations linking to web documents. Notably, these statements are supported by one or more citations. For this benchmark, human assessors are enrolled to annotate each statement-citation pair based on the degree to which the citation supports the associated statement.

\paragraph{Data Statistics}

The \gense{} dataset comprises a total of $12,681$ statement-citation pairs. For each pair, human assessors categorize the citation into one of three categories of support levels: full, partial, or no support. 
The details of data statistics are shown in Table \ref{tab:data_stats}.
Notably, for citations classified under the full or partial support categories, human assessors additionally extract explicit evidence sentences from the citation that support the associated statement.

\paragraph{Data Processing}

While the \gense{} dataset aligns well with our research objectives, we encounter a significant challenge: the extensive length of most citations within the dataset. 
These citations often comprise a web document with thousands of words, far exceeding the maximum input capacity of most faithfulness metrics, which is limited to $512$ tokens.
This limitation necessitates input truncation, potentially compromising the reliability of faithfulness metrics.
To mitigate this issue, we adopt a strategy similar to previous studies~\cite{zha-etal-2023-alignscore}. 
Specifically, we segment each cited document into shorter text chunks, with a maximum length of $150$ words per chunk. These text chunks, along with their corresponding statements, serve as the inputs for faithfulness metrics to predicted metric scores.
Furthermore, to determine human judgments for the text chunks, we employ the Jaccard similarity index to identify text chunks containing human-annotated evidence sentences, classifying them as either fully or partially supporting text chunks.

\subsection{Evaluation Metrics} 
\label{subsec:meta_eval}

We report Pearson, Spearman, and Kendall coefficients for correlation analysis, as recommended by previous research~\cite{pagnoni-etal-2021-understanding}.
In terms of classification evaluation, following previous studies~\cite{honovich-etal-2022-true-evaluating,ma-etal-2023-bump}, we report the macro-averaged Receiver Operating Characteristic-Area Under Curve (ROC-AUC) score, as it obviates the need for manual threshold setting for each binary classification task.
For retrieval evaluation, we report standard normalized discounted cumulative gain (NDCG@n) scores where $n \in \{5, 10, 20\}$.
\begin{table}[tbp]
\centering
\resizebox{0.985\linewidth}{!}{
\begin{tabular}{@{} lccc@{}}
\toprule[1pt]
\textbf{Metric}  & \textbf{Pearson} & \textbf{Spearman} & \textbf{Kendall}  \\
\midrule
\multicolumn{2}{@{} l}{\textit{LLM-based}} \\
\quad \gptcon{}   & 0.023 & 0.057 & 0.035 \\
\quad \gptdis{}   & 0.101 & 0.181 & 0.128 \\
\multicolumn{2}{@{} l}{\textit{Entailment-based}}  \\
\quad \factcc{}           & 0.121 & 0.199 & 0.140   \\
\quad \summaczsfull{}     & 0.364 & 0.180 & 0.137   \\
\quad \summacconvfull{}   & 0.565 & 0.444 & 0.342  \\
\quad \alignscore{}       & 0.585 & \underline{0.488} & \underline{0.393}   \\
\quad \autoais{}          & \textbf{0.638} & \textbf{0.639} & \textbf{0.547}   \\
\multicolumn{2}{@{} l}{\textit{Similarity-based}}  \\
\quad \bertscore{}        & 0.542 & 0.227 & 0.170   \\
\quad \bartscore{}        & \underline{0.598} & 0.235 & 0.176 \\
\bottomrule[1pt]
\end{tabular}
}
\caption{Correlation coefficients between human-annotated support levels and metric scores on the \gense{} dataset. The best and second-best correlations are marked in \textbf{bold} and \underline{underline}, respectively.}
\label{tab:correlation_results}
\end{table}

\begin{table*}[t]
\centering
\begin{tabular}{@{}llcccc@{}}
\toprule[1pt]
\textbf{Category} & \textbf{Metric}  & \textbf{\isfullno{}} & \textbf{\ispartialfull{}} & \textbf{\ispartialno{}} & \textbf{Overall}  \\
\midrule

\multirow{2}{*}{LLM-based} & \gptcon{} & 54.80 & 54.13 & 51.60 & 53.51 \\
& \gptdis{} & 57.84 & 52.79 & 55.48 & 55.37 \\
\cdashline{1-6}[1pt/1.5pt]
\multirow{5}{*}{Entailment-based} & \factcc{}              & 68.45 & 62.58 & 56.39 & 62.47 \\
& \summaczsfull{}     & 78.60 & 72.96 & 58.67 & 70.08 \\ 
& \summacconvfull{}   & 85.01 & 78.74 & 61.84 & 75.20 \\
& \alignscore{}       & 90.79 & \underline{81.41} & 69.78 & 80.66 \\
& \autoais{}          & \textbf{92.61} & \textbf{82.31} & \underline{73.90} & \textbf{82.94}  \\ 
\cdashline{1-6}[1pt/1.5pt]
\multirow{2}{*}{Similarity-based} & \bartscore{}           & 87.43 & 75.42 & 71.34 & 78.07 \\
& \bertscore{}           & \underline{91.55} & 75.94 & \textbf{78.72} & \underline{82.07} \\
\bottomrule[1pt]
\end{tabular}
\caption{Classification performance of faithfulness metrics regarding ROC-AUC score (\%) on the \gense{} dataset. The overall performance is the macro-averaged performance of three binary classification settings. The best and second-best scores are marked in \textbf{bold} and \underline{underline}, respectively.}
\label{tab:classification_results}
\end{table*}

\section{Results and Analyses}
\label{subsec:results}

In this section, we discuss the performance of faithfulness metrics across three distinct evaluation protocols. Subsequently, we conduct a qualitative analysis through case studies.

\subsection{Correlation Results} 

The correlation results are demonstrated in Table \ref{tab:correlation_results}. The following observations can be made:
\begin{enumerate*}[label=\arabic*)]
    \item The best-performing metrics reveal moderate correlations when analyzed using the Pearson coefficient. For instance, \autoais{} achieves the highest Pearson coefficient, recording a value of $0.638$, largely surpassing the second-best \bartscore{}, which posts a coefficient of $0.598$.
    
    \item 
    There is notable variation in correlation trends among high-performing metrics. \bartscore{} shows the second-best Pearson correlation but much lower Spearman and Kendall correlations. This divergence likely arises from the Pearson coefficient's assumption of linear relationships between two variables, which is often invalid in automated citation evaluation.
    
    \item Similarity-based metrics generally show lower Spearman and Kendall correlations compared to Pearson. For instance, \bertscore{} has a substantial Pearson correlation of $0.542$ but lower Spearman and Kendall correlations of $0.227$ and $0.170$. This indicates that similarity-based metrics do not align well with human judgments, highlighting their limitations in fine-grained support scenarios.
    
    \item LLM-based metrics show little correlation with human judgments among all correlation coefficients, with the correlation of the \gptcon{} metric being almost zero. This finding suggests a negligible relationship between LLM-based metric scores and human judgments. Furthermore, the \gptdis{} metric significantly outperforms \gptcon{}, highlighting that more fine-grained support levels present greater challenges in correlation analysis.
\end{enumerate*}

\begin{figure*}[t]
    \centering
    \includegraphics[trim=0cm 0cm 0cm 0.0cm,width=0.985\linewidth
]{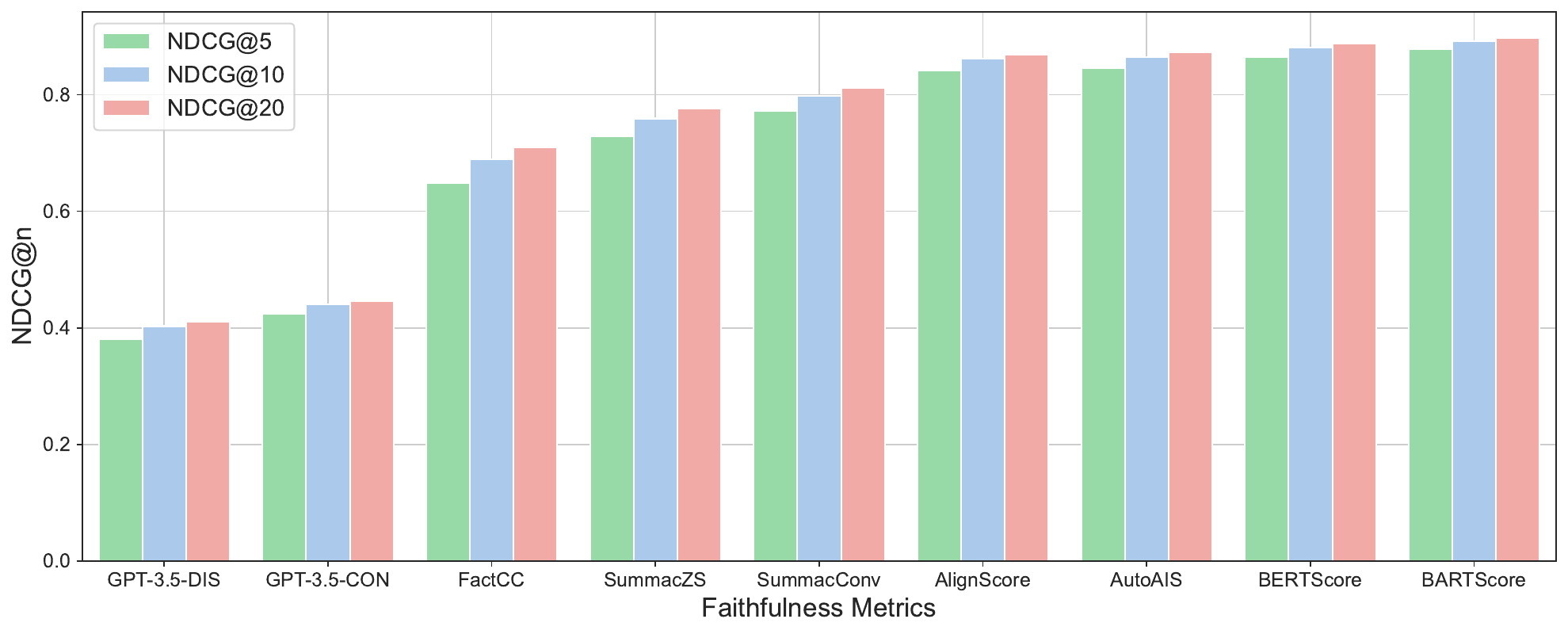}
    \caption{Retrieval performance of faithfulness metrics regarding NDCG@n scores on the \gense{} dataset. Note that we assign relevance labels $2$, $1$, and $0$ to full, partial, and no support, respectively (shown in the color). 
    }
    \label{fig:retrieval_hist}
\end{figure*}

\subsection{Classification Results}

Table~\ref{tab:classification_results} presents the results of the classification evaluation.  The observations can be summarized as follows:
\begin{enumerate*}[label=\arabic*)]
    \item Among all three binary classification task settings, most faithfulness metrics demonstrate superior performance in the \isfullno{} setting. Notably, entailment-based \autoais{} achieves the highest ROC-AUC score of $92.61$, which shows significant discriminability between full support and no support instances. This can be attributed to its much more extensive parameters compared to other entailment-based metrics.
    
    \item We observe the performance decline across the other two settings (i.e. FS-vs-PS and PS-vs-NS). For instance, when comparing the \isfullno{} and \ispartialno{} settings, the ROC-AUC score of \autoais{} diminishes from $92.61$ to $73.90$. This decline indicates that even the best-performing metric struggles with granular sensitivity to varying levels of support.
    
    \item While entailment-based \autoais{} generally surpasses other metrics, it is outperformed by similarity-based \bertscore{} in the \ispartialno{} setting. Interestingly, while most metrics perform worst in this setting, \bertscore{} shows its least effectiveness in \ispartialfull{}. This highlights the unique prediction behaviors of different metrics across binary classification settings.
    
    \item The performance of LLM-based metrics significantly lags behind other metrics. For instance, \gptdis{} achieves only a ROC-AUC score of $57.84$ in the \isfullno{} setting, markedly lower than the best-performing \autoais{}, which achieves a ROC-AUC score of $92.61$. Furthermore, the overall performance of LLM-based metrics approaches random guessing. This underscores the inefficacy of LLM-based metrics in distinguishing fine-grained support levels.
\end{enumerate*}

\subsection{Retrieval Results}

Figure~\ref{fig:retrieval_hist} presents the results of the retrieval evaluation. The key findings are as follows:
\begin{enumerate*}[label=\arabic*)]
    \item Similarity-based metrics, \bartscore{} and \bertscore{}, outperform other entailment-based metrics in all NDCG@n scores. For instance, entailment-based \autoais{} exhibits weaker NDCG@5 scores than \bartscore{}. This is likely because entailment-based metrics are more sensitive to noisy information than similarity-based metrics, as many irrelevant documents exist in retrieval scenarios. It suggests the need for the robustness improvements of metrics in post-hoc retrieval scenarios.  %
    
    \item The best-performing \bertscore{} achieves more than twice the NDCG@n scores compared to LLM-based metrics. This result suggests that LLM-based metrics are ineffective in ranking documents with higher support levels. A plausible explanation is that LLM-based metrics lack fine-grained sensitivity to variations in support levels. Interestingly, our observations reveal that \gptcon{} surpasses \gptdis{}, highlighting the advantage of fine-grained scoring methods in retrieval evaluation.
    
    \item NDCG@n scores effectively capture the performance variations as the number of text chunks increases. For instance, as the chunk count increases, \bartscore{} shows a marginal performance improvement, while \factcc{} exhibits a more pronounced enhancement. %
\end{enumerate*}

\begin{table*}[t]
\renewcommand\arraystretch{1.2}
\small
\centering
\resizebox{0.985\linewidth}{!}{
\begin{tabular}{p{0.2\textwidth}<{\raggedright}p{0.8\textwidth}}
\toprule[1pt]
\textbf{Error Reason} & \textbf{Example} \\
\midrule[1pt]

The citation does not explicitly mention coreference. &

\textbf{Statement:} Others believe that \colorbox{mygreen}{performance-enhancing drugs} should be allowed in sports.

\textbf{Citation:} However, if children are allowed to train as professional athletes, then they should be allowed to take \colorbox{mygreen}{the same drugs}, provided that they are no more dangerous than their training is \ldots

\textbf{Human Judgment:} full support

\textbf{Metric Score:}  0.055 (no support) \\
\hline

The complex statement includes independent claims. & 

\textbf{Statement:} \colorbox{mygreen}{Love leads to growth} \colorbox{myred}{while being in love is about ownership}  \ldots

\textbf{Citation:} ``\colorbox{mygreen}{Growing to love} the real person and accepting who they are, with both strengths and weaknesses, can make a wonderful difference in your relationship,'' McCoy says \ldots

\textbf{Human Judgment:} partial support

\textbf{Metric Score:}  0.0004 (no support) \\

\hline

The citation is semantically similar but non-supporting. & 

\textbf{Statement:} Carpal tunnel syndrome can be treated with various methods, including \colorbox{myred}{wrist splinting, anti-inflammatory medication}, and surgery.

\textbf{Citation:} If diagnosed and treated early, the symptoms of carpal tunnel syndrome can often be relieved \colorbox{myred}{without surgery}. If your diagnosis is uncertain or if your symptoms are mild, your doctor will recommend \colorbox{myred}{nonsurgical treatment} first \ldots

\textbf{Human Judgment:} no support

\textbf{Metric Score:}  0.52 (partial support) \\

\hline

\end{tabular}
}
\caption{Case study of the faithfulness metric \autoais{}. \colorbox{mygreen}{Green phrases} indicate supported content in the statement and corresponding supporting evidence. \colorbox{myred}{Red phrases} indicate unsupported content in the statement and corresponding misleading information in the citation.}
\label{tab:case_study}
\end{table*}

\subsection{Case Study}
\label{subsec:case_study}

Table~\ref{tab:case_study} presents three cases of \autoais{}.
In the first example, where human judgment indicates full support. \autoais{} incorrectly assigns a very low score. This may be due to the lack of explicit mention of drug coreference in the cited text chunk. This indicates coreference resolution remains a significant challenge in automated citation evaluation.
In the second example, where human judgment indicates partial support. The complex statement implicitly contains two independent claims that require verification. However, the provided citation fails to offer sufficient evidence, resulting in an almost zero metric score.
In the third example, where human judgment indicates no support. The given citation is semantically similar to the statement, leading to a metric score of partial support. Despite this semantic similarity, specific treatments mentioned in the statement, such as wrist splinting, are not explicitly referenced in the citation.

\section{Discussions}

Overall, our results across three evaluation protocols indicate that the evaluation protocols are complementary and should be integrated for a comprehensive assessment of metrics.
Based on the evaluation results, we further propose the following practical recommendations to develop more effective metrics for automated citation evaluation:
\begin{enumerate*}[label=\arabic*)]
    \item \textbf{Development of training resources:} motivated by the observation that the best-performing metrics still struggle with identifying partial support, we recommend the development of training resources that include fine-grained support level annotations. These resources could significantly enhance the metrics' fine-grained sensitivity to varying support levels;
    \item \textbf{Introduction of contrastive learning:} to improve the robustness of metrics in post-hoc retrieval scenarios, we recommend fine-tuning metrics using contrastive learning frameworks. This method has demonstrated effectiveness across various information retrieval tasks~\cite{izacard2022contriever}.
    \item \textbf{Development of more explainable metrics:} traditional faithfulness metrics often only provide final scores without sufficient explainability~\cite{xu-etal-2023-instructscore}. This limitation hinders a deeper understanding of the models' behavior. Therefore, it is crucial to develop more explainable faithfulness metrics, potentially using large language models (LLMs). 
\end{enumerate*}

\section{Conclusion}

We propose a comparative evaluation framework to explore the efficacy of faithfulness metrics beyond the binary scenario by examining three levels of citation support. Our framework employs correlation analysis, classification evaluation, and retrieval evaluation to measure the alignment between metric scores and human judgments. 
Experimental results reveal that no single metric consistently excels across all evaluation protocols, indicating the complexity of automated citation evaluation and the limitations of existing faithfulness metrics. We provide practical suggestions based on the findings.

\section*{Limitations}

In this work, we consider a citation that explicitly contains human-annotated evidence as the fully supporting citation for each statement. However, for some complex statements, their evidence is distributed among multiple citations. For instance, about $2\%$ statements on the \gense{} dataset require multiple citations to be fully supported.
Also, we focus on statement-level citation evaluation. Since answer-level citation evaluation is much more complicated and requires proper aggregation methods, we leave this exploration as future work.
We do not evaluate QA-based faithfulness metrics as a recent study shows that such metrics have some fundamental issues, such as failing to localize errors~\cite{kamoi-etal-2023-shortcomings}. However different findings could be explored with QA-based metrics.
Finally, there is another line of work, in which they break down a complex statement into multiple atomic facts.

\section*{Ethical Considerations}

We realized there are some risks in exploring citation evaluation for LLM-generated text. Since we have used publicly available datasets and open-source implementation of faithfulness metrics, we carefully avoid potential ethical problems caused by datasets or open-source codes. As we address the issue of the effectiveness of faithfulness metrics for LLM-generated text, concerning hallucination. We acknowledged the hallucinated text generated by LLMs may contain potentially harmful or misleading information. Our final goal is to mitigate such hallucination issues, which should support the discussion around hallucinations of LLMs and all ethical aspects around them.

\section*{Acknowledgments}

This research was supported by the China Scholarship Council
(CSC) under grant number 202008440470. 
The views expressed in the content belong solely to the authors and may not reflect the perspectives or endorsements of their respective employers or sponsors.

\bibliography{main}

\begin{thebibliography}{66}
\providecommand{\natexlab}[1]{#1}

\bibitem[{Asai et~al.(2024)Asai, Wu, Wang, Sil, and Hajishirzi}]{asai2024selfrag}
Akari Asai, Zeqiu Wu, Yizhong Wang, Avirup Sil, and Hannaneh Hajishirzi. 2024.
\newblock \href {https://openreview.net/forum?id=hSyW5go0v8} {Self-rag: Learning to retrieve, generate, and critique through self-reflection}.
\newblock In \emph{The Twelfth International Conference on Learning Representations, {ICLR} 2024, Vienna, Austria, May 7-11, 2024}.

\bibitem[{Bohnet et~al.(2022)Bohnet, Tran, Verga, Aharoni, Andor, Soares, Ciaramita, Eisenstein, Ganchev, Herzig et~al.}]{bohnet2022attributed}
Bernd Bohnet, Vinh~Q Tran, Pat Verga, Roee Aharoni, Daniel Andor, Livio~Baldini Soares, Massimiliano Ciaramita, Jacob Eisenstein, Kuzman Ganchev, Jonathan Herzig, et~al. 2022.
\newblock Attributed question answering: Evaluation and modeling for attributed large language models.
\newblock \emph{arXiv preprint arXiv:2212.08037}.

\bibitem[{Devlin et~al.(2019)Devlin, Chang, Lee, and Toutanova}]{devlin2019bert}
Jacob Devlin, Ming-Wei Chang, Kenton Lee, and Kristina Toutanova. 2019.
\newblock \href {https://doi.org/10.18653/v1/n19-1423} {{{BERT}}: {{Pre-training}} of deep bidirectional transformers for language understanding}.
\newblock In \emph{Proceedings of the 2019 Conference of the North American Chapter of the Association for Computational Linguistics: {{Human}} Language Technologies, {{NAACL-HLT}} 2019, Minneapolis, {{MN}}, {{USA}}, June 2-7, 2019, Volume 1 (Long and Short Papers)}, pages 4171--4186.

\bibitem[{Durmus et~al.(2020)Durmus, He, and Diab}]{durmus-etal-2020-feqa}
Esin Durmus, He~He, and Mona Diab. 2020.
\newblock \href {https://doi.org/10.18653/v1/2020.acl-main.454} {{{FEQA}}: {{A}} question answering evaluation framework for faithfulness assessment in abstractive summarization}.
\newblock In \emph{Proceedings of the 58th Annual Meeting of the Association for Computational Linguistics}, pages 5055--5070.

\bibitem[{Fabbri et~al.(2022)Fabbri, Wu, Liu, and Xiong}]{fabbri-etal-2022-qafacteval}
Alexander Fabbri, Chien-Sheng Wu, Wenhao Liu, and Caiming Xiong. 2022.
\newblock \href {https://doi.org/10.18653/v1/2022.naacl-main.187} {{{QAFactEval}}: {{Improved QA-Based}} factual consistency evaluation for summarization}.
\newblock In \emph{Proceedings of the 2022 Conference of the North American Chapter of the Association for Computational Linguistics: {{Human}} Language Technologies}, pages 2587--2601.

\bibitem[{Falke et~al.(2019)Falke, Ribeiro, Utama, Dagan, and Gurevych}]{falke-etal-2019-ranking}
Tobias Falke, Leonardo F.~R. Ribeiro, Prasetya~Ajie Utama, Ido Dagan, and Iryna Gurevych. 2019.
\newblock \href {https://doi.org/10.18653/v1/P19-1213} {Ranking generated summaries by correctness: {{An}} interesting but challenging application for natural language inference}.
\newblock In \emph{Proceedings of the 57th Annual Meeting of the Association for Computational Linguistics}, pages 2214--2220.

\bibitem[{Fan et~al.(2019)Fan, Jernite, Perez, Grangier, Weston, and Auli}]{fan2019eli5}
Angela Fan, Yacine Jernite, Ethan Perez, David Grangier, Jason Weston, and Michael Auli. 2019.
\newblock \href {https://doi.org/10.18653/v1/p19-1346} {{{ELI5}}: {{Long Form Question Answering}}}.
\newblock In \emph{Proceedings of the 57th {{Conference}} of the {{Association}} for {{Computational Linguistics}}, {{ACL}} 2019, {{Florence}}, {{Italy}}, {{July}} 28- {{August}} 2, 2019, {{Volume}} 1: {{Long Papers}}}, pages 3558--3567.

\bibitem[{Gao et~al.(2023{\natexlab{a}})Gao, Dai, Pasupat, Chen, Chaganty, Fan, Zhao, Lao, Lee, Juan, and Guu}]{gao-etal-2023-rarr}
Luyu Gao, Zhuyun Dai, Panupong Pasupat, Anthony Chen, Arun~Tejasvi Chaganty, Yicheng Fan, Vincent Zhao, Ni~Lao, Hongrae Lee, Da-Cheng Juan, and Kelvin Guu. 2023{\natexlab{a}}.
\newblock {{RARR}}: {{Researching}} and revising what language models say, using language models.
\newblock In \emph{Proceedings of the 61st Annual Meeting of the Association for Computational Linguistics (Volume 1: {{Long}} Papers)}, pages 16477--16508.

\bibitem[{Gao et~al.(2023{\natexlab{b}})Gao, Yen, Yu, and Chen}]{gao-etal-2023-enabling}
Tianyu Gao, Howard Yen, Jiatong Yu, and Danqi Chen. 2023{\natexlab{b}}.
\newblock \href {https://doi.org/10.18653/v1/2023.emnlp-main.398} {Enabling large language models to generate text with citations}.
\newblock In \emph{Proceedings of the 2023 Conference on Empirical Methods in Natural Language Processing}, pages 6465--6488.

\bibitem[{Gou et~al.(2024)Gou, Shao, Gong, Shen, Yang, Duan, and Chen}]{gou2024critic}
Zhibin Gou, Zhihong Shao, Yeyun Gong, Yelong Shen, Yujiu Yang, Nan Duan, and Weizhu Chen. 2024.
\newblock \href {https://openreview.net/forum?id=Sx038qxjek} {{CRITIC:} large language models can self-correct with tool-interactive critiquing}.
\newblock In \emph{The Twelfth International Conference on Learning Representations, {ICLR} 2024, Vienna, Austria, May 7-11, 2024}.

\bibitem[{He et~al.(2021)He, Liu, Gao, and Chen}]{he2021deberta}
Pengcheng He, Xiaodong Liu, Jianfeng Gao, and Weizhu Chen. 2021.
\newblock \href {https://openreview.net/forum?id=XPZIaotutsD} {Deberta: Decoding-{{Enhanced Bert}} with {{Disentangled Attention}}}.
\newblock In \emph{9th {{International Conference}} on {{Learning Representations}}, {{ICLR}} 2021, {{Virtual Event}}, {{Austria}}, {{May}} 3-7, 2021}.

\bibitem[{Hermann et~al.(2015)Hermann, Kociský, Grefenstette, Espeholt, Kay, Suleyman, and Blunsom}]{hermann2015cnndm}
Karl~Moritz Hermann, Tomás Kociský, Edward Grefenstette, Lasse Espeholt, Will Kay, Mustafa Suleyman, and Phil Blunsom. 2015.
\newblock \href {https://proceedings.neurips.cc/paper/2015/hash/afdec7005cc9f14302cd0474fd0f3c96-Abstract.html} {Teaching machines to read and comprehend}.
\newblock In \emph{Advances in Neural Information Processing Systems 28: {{Annual}} Conference on Neural Information Processing Systems 2015, December 7-12, 2015, Montreal, Quebec, Canada}, pages 1693--1701.

\bibitem[{Honovich et~al.(2022)Honovich, Aharoni, Herzig, Taitelbaum, Kukliansy, Cohen, Scialom, Szpektor, Hassidim, and Matias}]{honovich-etal-2022-true-evaluating}
Or~Honovich, Roee Aharoni, Jonathan Herzig, Hagai Taitelbaum, Doron Kukliansy, Vered Cohen, Thomas Scialom, Idan Szpektor, Avinatan Hassidim, and Yossi Matias. 2022.
\newblock \href {https://doi.org/10.18653/v1/2022.naacl-main.287} {{{TRUE}}: {{Re-evaluating}} factual consistency evaluation}.
\newblock In \emph{Proceedings of the 2022 Conference of the North American Chapter of the Association for Computational Linguistics: {{Human}} Language Technologies}, pages 3905--3920.

\bibitem[{Huang et~al.(2024{\natexlab{a}})Huang, Wu, Hu, and Wang}]{huang2024traincitation}
Chengyu Huang, Zeqiu Wu, Yushi Hu, and Wenya Wang. 2024{\natexlab{a}}.
\newblock \href {https://arxiv.org/abs/2402.04315} {Training language models to generate text with citations via fine-grained rewards}.
\newblock \emph{CoRR}, abs/2402.04315.

\bibitem[{Huang et~al.(2019)Huang, Dao, Alfadly, and Ghanem}]{huang2019novel}
Jia-Hong Huang, Cuong~Duc Dao, Modar Alfadly, and Bernard Ghanem. 2019.
\newblock A novel framework for robustness analysis of visual qa models.
\newblock In \emph{Proceedings of the AAAI Conference on Artificial Intelligence}, volume~33, pages 8449--8456.

\bibitem[{Huang et~al.(2021{\natexlab{a}})Huang, Murn, Mrak, and Worring}]{huang2021gpt2mvs}
Jia-Hong Huang, Luka Murn, Marta Mrak, and Marcel Worring. 2021{\natexlab{a}}.
\newblock Gpt2mvs: Generative pre-trained transformer-2 for multi-modal video summarization.
\newblock In \emph{Proceedings of the International Conference on Multimedia Retrieval}, pages 580--589.

\bibitem[{Huang and Worring(2020)}]{huang2020query}
Jia-Hong Huang and Marcel Worring. 2020.
\newblock Query-controllable video summarization.
\newblock In \emph{Proceedings of the International Conference on Multimedia Retrieval}, pages 242--250.

\bibitem[{Huang et~al.(2021{\natexlab{b}})Huang, Yang, Liu, Tian, Liu, Wu, Lin, Wang, Morikawa, Chang et~al.}]{huang2021deepopht}
Jia-Hong Huang, C-H~Huck Yang, Fangyu Liu, Meng Tian, Yi-Chieh Liu, Ting-Wei Wu, I~Lin, Kang Wang, Hiromasa Morikawa, Hernghua Chang, et~al. 2021{\natexlab{b}}.
\newblock Deepopht: medical report generation for retinal images via deep models and visual explanation.
\newblock In \emph{Proceedings of the IEEE/CVF winter conference on applications of computer vision}, pages 2442--2452.

\bibitem[{Huang et~al.(2024{\natexlab{b}})Huang, Yang, Shen, Pacces, and Kanoulas}]{huang2024optimizing}
Jia-Hong Huang, Chao-Chun Yang, Yixian Shen, Alessio~M Pacces, and Evangelos Kanoulas. 2024{\natexlab{b}}.
\newblock Optimizing numerical estimation and operational efficiency in the legal domain through large language models.
\newblock In \emph{ACM International Conference on Information and Knowledge Management (CIKM)}.

\bibitem[{Huang et~al.(2022)Huang, Yang, Chen, Brown, and Worring}]{huang2022causal}
Jia-Hong Huang, Chao-Han~Huck Yang, Pin-Yu Chen, Andrew Brown, and Marcel Worring. 2022.
\newblock Causal video summarizer for video exploration.
\newblock In \emph{2022 IEEE International Conference on Multimedia and Expo (ICME)}, pages 1--6. IEEE.

\bibitem[{Huang et~al.(2023)Huang, Yang, Chen, Chen, and Worring}]{huang2023causalainer}
Jia-Hong Huang, Chao-Han~Huck Yang, Pin-Yu Chen, Min-Hung Chen, and Marcel Worring. 2023.
\newblock Causalainer: Causal explainer for automatic video summarization.
\newblock In \emph{Proceedings of the IEEE/CVF Conference on Computer Vision and Pattern Recognition}, pages 2629--2635.

\bibitem[{Huang et~al.(2024{\natexlab{c}})Huang, Zhu, Shen, Rudinac, Pacces, and Kanoulas}]{huang2024novel}
Jia-Hong Huang, Hongyi Zhu, Yixian Shen, Stevan Rudinac, Alessio~M. Pacces, and Evangelos Kanoulas. 2024{\natexlab{c}}.
\newblock A novel evaluation framework for image2text generation.
\newblock In \emph{International ACM SIGIR Conference on Research and Development in Information Retrieval, LLM4Eval Workshop}.

\bibitem[{Huang and Chang(2023)}]{huang2023citation}
Jie Huang and Kevin Chen-Chuan Chang. 2023.
\newblock \href {https://doi.org/10.48550/arXiv.2307.02185} {Citation: {{A}} key to building responsible and accountable large language models}.
\newblock \emph{CoRR}, abs/2307.02185.

\bibitem[{Huang et~al.(2024{\natexlab{d}})Huang, Feng, Ma, Gu, Zhong, Feng, Yu, Peng, Tang, Tu, and Qin}]{huang-etal-2024-learning}
Lei Huang, Xiaocheng Feng, Weitao Ma, Yuxuan Gu, Weihong Zhong, Xiachong Feng, Weijiang Yu, Weihua Peng, Duyu Tang, Dandan Tu, and Bing Qin. 2024{\natexlab{d}}.
\newblock Learning fine-grained grounded citations for attributed large language models.
\newblock In \emph{Findings of the Association for Computational Linguistics {{ACL}} 2024}, pages 14095--14113.

\bibitem[{Izacard et~al.(2022)Izacard, Caron, Hosseini, Riedel, Bojanowski, Joulin, and Grave}]{izacard2022contriever}
Gautier Izacard, Mathilde Caron, Lucas Hosseini, Sebastian Riedel, Piotr Bojanowski, Armand Joulin, and Edouard Grave. 2022.
\newblock Unsupervised dense information retrieval with contrastive learning.
\newblock \emph{Transactions on Machine Learning Research}, 2022.

\bibitem[{Ji et~al.(2022)Ji, Lee, Frieske, Yu, Su, Xu, Ishii, Bang, Madotto, and Fung}]{ji2022hallucination}
Ziwei Ji, Nayeon Lee, Rita Frieske, Tiezheng Yu, Dan Su, Yan Xu, Etsuko Ishii, Yejin Bang, Andrea Madotto, and Pascale Fung. 2022.
\newblock Survey of hallucination in natural language generation.
\newblock \emph{ACM Computing Surveys}.

\bibitem[{Kamoi et~al.(2023)Kamoi, Goyal, and Durrett}]{kamoi-etal-2023-shortcomings}
Ryo Kamoi, Tanya Goyal, and Greg Durrett. 2023.
\newblock Shortcomings of question answering based factuality frameworks for error localization.
\newblock In \emph{Proceedings of the 17th {{Conference}} of the {{European Chapter}} of the {{Association}} for {{Computational Linguistics}}}, pages 132--146.

\bibitem[{Kang et~al.(2023)Kang, Ni, and Yao}]{kang2023ever}
Haoqiang Kang, Juntong Ni, and Huaxiu Yao. 2023.
\newblock \href {https://doi.org/10.48550/ARXIV.2311.09114} {Ever: {{Mitigating}} hallucination in large language models through real-time verification and rectification}.
\newblock \emph{CoRR}, abs/2311.09114.

\bibitem[{Karpukhin et~al.(2020)Karpukhin, Oguz, Min, Lewis, Wu, Edunov, Chen, and Yih}]{karpukhin-etal-2020-dense}
Vladimir Karpukhin, Barlas Oguz, Sewon Min, Patrick Lewis, Ledell Wu, Sergey Edunov, Danqi Chen, and Wen-tau Yih. 2020.
\newblock Dense passage retrieval for open-domain question answering.
\newblock In \emph{Proceedings of the 2020 Conference on Empirical Methods in Natural Language Processing ({{EMNLP}})}, pages 6769--6781.

\bibitem[{Kojima et~al.(2022)Kojima, Gu, Reid, Matsuo, and Iwasawa}]{kojima2022zerocot}
Takeshi Kojima, Shixiang~Shane Gu, Machel Reid, Yutaka Matsuo, and Yusuke Iwasawa. 2022.
\newblock Large language models are zero-shot reasoners.
\newblock In \emph{{{NeurIPS}}}.

\bibitem[{Kryscinski et~al.(2020)Kryscinski, McCann, Xiong, and Socher}]{kryscinski-etal-2020-evaluating}
Wojciech Kryscinski, Bryan McCann, Caiming Xiong, and Richard Socher. 2020.
\newblock \href {https://doi.org/10.18653/v1/2020.emnlp-main.750} {Evaluating the factual consistency of abstractive text summarization}.
\newblock In \emph{Proceedings of the 2020 Conference on Empirical Methods in Natural Language Processing ({{EMNLP}})}, pages 9332--9346.

\bibitem[{Kwiatkowski et~al.(2019)Kwiatkowski, Palomaki, and et~al.}]{kwiatkowski-etal-2019-natural}
Tom Kwiatkowski, Jennimaria Palomaki, and et~al. 2019.
\newblock \href {https://doi.org/10.1162/tacl_a_00276} {Natural questions: {{A}} benchmark for question answering research}.
\newblock \emph{Transactions of the Association for Computational Linguistics}, 7:452--466.

\bibitem[{Laban et~al.(2022)Laban, Schnabel, Bennett, and Hearst}]{laban-etal-2022-summac}
Philippe Laban, Tobias Schnabel, Paul~N. Bennett, and Marti~A. Hearst. 2022.
\newblock \href {https://doi.org/10.1162/tacl_a_00453} {{{SummaC}}: {{Re-visiting NLI-based}} models for inconsistency detection in summarization}.
\newblock \emph{Transactions of the Association for Computational Linguistics}, 10:163--177.

\bibitem[{Lewis et~al.(2020)Lewis, Liu, Goyal, Ghazvininejad, Mohamed, Levy, Stoyanov, and Zettlemoyer}]{lewis2020bart}
Mike Lewis, Yinhan Liu, Naman Goyal, Marjan Ghazvininejad, Abdelrahman Mohamed, Omer Levy, Veselin Stoyanov, and Luke Zettlemoyer. 2020.
\newblock \href {https://doi.org/10.18653/v1/2020.acl-main.703} {{{BART}}: {{Denoising}} sequence-to-sequence pre-training for natural language generation, translation, and comprehension}.
\newblock In \emph{Proceedings of the 58th Annual Meeting of the Association for Computational Linguistics, {{ACL}} 2020, Online, July 5-10, 2020}, pages 7871--7880.

\bibitem[{Li et~al.(2024{\natexlab{a}})Li, Sun, Hu, Liu, Hu, Liu, and Zhang}]{li-etal-2024-improving-attributed}
Dongfang Li, Zetian Sun, Baotian Hu, Zhenyu Liu, Xinshuo Hu, Xuebo Liu, and Min Zhang. 2024{\natexlab{a}}.
\newblock Improving attributed text generation of large language models via preference learning.
\newblock In \emph{Findings of the Association for Computational Linguistics {{ACL}} 2024}, pages 5079--5101.

\bibitem[{Li et~al.(2022)Li, Wu, Chen, Liu, Xiao, and Wu}]{li2022faithfulness}
Wei Li, Wenhao Wu, Moye Chen, Jiachen Liu, Xinyan Xiao, and Hua Wu. 2022.
\newblock Faithfulness in natural language generation: A systematic survey of analysis, evaluation and optimization methods.
\newblock \emph{arXiv preprint arXiv:2203.05227}.

\bibitem[{Li et~al.(2024{\natexlab{b}})Li, Zhu, Li, Yin, Sun, and Qiu}]{li2023llatrieval}
Xiaonan Li, Changtai Zhu, Linyang Li, Zhangyue Yin, Tianxiang Sun, and Xipeng Qiu. 2024{\natexlab{b}}.
\newblock \href {https://doi.org/10.18653/v1/2024.naacl-long.305} {{{LLatrieval}}: {{LLM-verified}} retrieval for verifiable generation}.
\newblock In \emph{Proceedings of the 2024 Conference of the North American Chapter of the Association for Computational Linguistics: {{Human}} Language Technologies (Volume 1: {{Long}} Papers)}, pages 5453--5471.

\bibitem[{Li et~al.(2024{\natexlab{c}})Li, Cao, Pan, Ma, and Sun}]{li2023towards}
Xinze Li, Yixin Cao, Liangming Pan, Yubo Ma, and Aixin Sun. 2024{\natexlab{c}}.
\newblock Towards verifiable generation: A benchmark for knowledge-aware language model attribution.
\newblock In \emph{Findings of the Association for Computational Linguistics {{ACL}} 2024}, pages 493--516.

\bibitem[{Li et~al.(2024{\natexlab{d}})Li, Yue, Liao, and Sun}]{li2024attrbench}
Yifei Li, Xiang Yue, Zeyi Liao, and Huan Sun. 2024{\natexlab{d}}.
\newblock {{AttributionBench}}: {{How}} hard is automatic attribution evaluation?
\newblock In \emph{Findings of the Association for Computational Linguistics {{ACL}} 2024}, pages 14919--14935.

\bibitem[{Liu et~al.(2023)Liu, Zhang, and Liang}]{liu-etal-2023-evaluating}
Nelson Liu, Tianyi Zhang, and Percy Liang. 2023.
\newblock \href {https://doi.org/10.18653/v1/2023.findings-emnlp.467} {Evaluating verifiability in generative search engines}.
\newblock In \emph{Findings of the Association for Computational Linguistics: {{EMNLP}} 2023}, pages 7001--7025.

\bibitem[{Liu et~al.(2019)Liu, Ott, Goyal, Du, Joshi, Chen, Levy, Lewis, Zettlemoyer, and Stoyanov}]{liu2019roberta}
Yinhan Liu, Myle Ott, Naman Goyal, Jingfei Du, Mandar Joshi, Danqi Chen, Omer Levy, Mike Lewis, Luke Zettlemoyer, and Veselin Stoyanov. 2019.
\newblock Roberta: A robustly optimized bert pretraining approach.
\newblock \emph{arXiv preprint arXiv:1907.11692}.

\bibitem[{Ma et~al.(2023)Ma, Cao, Logan~IV, Lu, Ran, Zhang, Tetreault, and Jaimes}]{ma-etal-2023-bump}
Liang Ma, Shuyang Cao, Robert~L Logan~IV, Di~Lu, Shihao Ran, Ke~Zhang, Joel Tetreault, and Alejandro Jaimes. 2023.
\newblock {{BUMP}}: {{A}} benchmark of unfaithful minimal pairs for meta-evaluation of faithfulness metrics.
\newblock In \emph{Proceedings of the 61st Annual Meeting of the Association for Computational Linguistics (Volume 1: {{Long}} Papers)}, pages 12788--12812.

\bibitem[{Maynez et~al.(2020)Maynez, Narayan, Bohnet, and McDonald}]{maynez-etal-2020-faithfulness}
Joshua Maynez, Shashi Narayan, Bernd Bohnet, and Ryan McDonald. 2020.
\newblock \href {https://doi.org/10.18653/v1/2020.acl-main.173} {On faithfulness and factuality in abstractive summarization}.
\newblock In \emph{Proceedings of the 58th Annual Meeting of the Association for Computational Linguistics}, pages 1906--1919.

\bibitem[{Pagnoni et~al.(2021)Pagnoni, Balachandran, and Tsvetkov}]{pagnoni-etal-2021-understanding}
Artidoro Pagnoni, Vidhisha Balachandran, and Yulia Tsvetkov. 2021.
\newblock \href {https://doi.org/10.18653/v1/2021.naacl-main.383} {Understanding factuality in abstractive summarization with {{FRANK}}: {{A}} benchmark for factuality metrics}.
\newblock In \emph{Proceedings of the 2021 Conference of the North American Chapter of the Association for Computational Linguistics: {{Human}} Language Technologies}, pages 4812--4829.

\bibitem[{Raffel et~al.(2020)Raffel, Shazeer, Roberts, and et~al}]{raffel2020exploring}
Colin Raffel, Noam Shazeer, Adam Roberts, and et~al. 2020.
\newblock \href {http://jmlr.org/papers/v21/20-074.html} {Exploring the limits of transfer learning with a unified text-to-text transformer}.
\newblock \emph{Journal of Machine Learning Research}, 21:140:1--140:67.

\bibitem[{Rashkin et~al.(2023)Rashkin, Nikolaev, Lamm, Aroyo, Collins, Das, Petrov, Tomar, Turc, and Reitter}]{rashkin2021attribution}
Hannah Rashkin, Vitaly Nikolaev, Matthew Lamm, Lora Aroyo, Michael Collins, Dipanjan Das, Slav Petrov, Gaurav~Singh Tomar, Iulia Turc, and David Reitter. 2023.
\newblock Measuring attribution in natural language generation models.
\newblock \emph{Computational Linguistics}, 49(4):777--840.

\bibitem[{Scialom et~al.(2021)Scialom, Dray, Lamprier, Piwowarski, Staiano, Wang, and Gallinari}]{scialom-etal-2021-questeval}
Thomas Scialom, Paul-Alexis Dray, Sylvain Lamprier, Benjamin Piwowarski, Jacopo Staiano, Alex Wang, and Patrick Gallinari. 2021.
\newblock \href {https://doi.org/10.18653/v1/2021.emnlp-main.529} {{{QuestEval}}: {{Summarization}} asks for fact-based evaluation}.
\newblock In \emph{Proceedings of the 2021 Conference on Empirical Methods in Natural Language Processing}, pages 6594--6604.

\bibitem[{Shen et~al.(2024)Shen, Zhou, Zhao, Chen, and Liu}]{shen2024citekit}
Jiajun Shen, Tong Zhou, Suifeng Zhao, Yubo Chen, and Kang Liu. 2024.
\newblock Citekit: A modular toolkit for large language model citation generation.
\newblock \emph{arXiv preprint arXiv:2408.04662}.

\bibitem[{Sun et~al.(2023)Sun, Cai, Wang, Hou, Wei, Wang, Zhang, and Yin}]{sun2023towards}
Hao Sun, Hengyi Cai, Bo~Wang, Yingyan Hou, Xiaochi Wei, Shuaiqiang Wang, Yan Zhang, and Dawei Yin. 2023.
\newblock \href {https://doi.org/10.48550/ARXIV.2312.09075} {Towards verifiable text generation with evolving memory and self-reflection}.
\newblock \emph{CoRR}, abs/2312.09075.

\bibitem[{Tahaei et~al.(2024)Tahaei, Jafari, Rashid, {Alfonso-Hermelo}, Bibi, Wu, Ghodsi, Chen, and Rezagholizadeh}]{tahaei-etal-2024-efficient}
Marzieh Tahaei, Aref Jafari, Ahmad Rashid, David {Alfonso-Hermelo}, Khalil Bibi, Yimeng Wu, Ali Ghodsi, Boxing Chen, and Mehdi Rezagholizadeh. 2024.
\newblock Efficient citer: {{Tuning}} large language models for enhanced answer quality and verification.
\newblock In \emph{Findings of the Association for Computational Linguistics: {{NAACL}} 2024}, pages 4443--4450.

\bibitem[{Wang et~al.(2020)Wang, Cho, and Lewis}]{wang-etal-2020-asking}
Alex Wang, Kyunghyun Cho, and Mike Lewis. 2020.
\newblock \href {https://doi.org/10.18653/v1/2020.acl-main.450} {Asking and answering questions to evaluate the factual consistency of summaries}.
\newblock In \emph{Proceedings of the 58th Annual Meeting of the Association for Computational Linguistics}, pages 5008--5020.

\bibitem[{Wei et~al.(2022)Wei, Wang, Schuurmans, Bosma, Ichter, Xia, Chi, Le, and Zhou}]{wei2022cot}
Jason Wei, Xuezhi Wang, Dale Schuurmans, Maarten Bosma, Brian Ichter, Fei Xia, Ed~H. Chi, Quoc~V. Le, and Denny Zhou. 2022.
\newblock Chain-of-thought prompting elicits reasoning in large language models.
\newblock In \emph{{{NeurIPS}}}.

\bibitem[{Xia et~al.(2024)Xia, Wang, Liang, Zhang, Zhou, Deng, Yu, and Xiao}]{xia2024ground}
Sirui Xia, Xintao Wang, Jiaqing Liang, Yifei Zhang, Weikang Zhou, Jiaji Deng, Fei Yu, and Yanghua Xiao. 2024.
\newblock Ground every sentence: Improving retrieval-augmented llms with interleaved reference-claim generation.
\newblock \emph{arXiv preprint arXiv:2407.01796}.

\bibitem[{Xu et~al.(2023)Xu, Wang, Pan, Song, Freitag, Wang, and Li}]{xu-etal-2023-instructscore}
Wenda Xu, Danqing Wang, Liangming Pan, Zhenqiao Song, Markus Freitag, William Wang, and Lei Li. 2023.
\newblock {{INSTRUCTSCORE}}: {{Towards}} explainable text generation evaluation with automatic feedback.
\newblock In \emph{Proceedings of the 2023 Conference on Empirical Methods in Natural Language Processing}, pages 5967--5994.

\bibitem[{Ye et~al.(2024)Ye, Sun, Arik, and Pfister}]{ye-etal-2024-effective}
Xi~Ye, Ruoxi Sun, Sercan Arik, and Tomas Pfister. 2024.
\newblock Effective large language model adaptation for improved grounding and citation generation.
\newblock In \emph{Proceedings of the 2024 Conference of the North American Chapter of the Association for Computational Linguistics: {{Human}} Language Technologies (Volume 1: {{Long}} Papers)}, pages 6237--6251.

\bibitem[{Yuan et~al.(2021)Yuan, Neubig, and Liu}]{yuan2021bartscore}
Weizhe Yuan, Graham Neubig, and Pengfei Liu. 2021.
\newblock \href {https://proceedings.neurips.cc/paper/2021/hash/e4d2b6e6fdeca3e60e0f1a62fee3d9dd-Abstract.html} {{{BARTScore}}: {{Evaluating Generated Text}} as {{Text Generation}}}.
\newblock In \emph{Advances in {{Neural Information Processing Systems}} 34: {{Annual Conference}} on {{Neural Information Processing Systems}} 2021, {{NeurIPS}} 2021, {{December}} 6-14, 2021, Virtual}, pages 27263--27277.

\bibitem[{Yue et~al.(2023)Yue, Wang, Chen, Zhang, Su, and Sun}]{yue-etal-2023-automatic}
Xiang Yue, Boshi Wang, Ziru Chen, Kai Zhang, Yu~Su, and Huan Sun. 2023.
\newblock \href {https://doi.org/10.18653/v1/2023.findings-emnlp.307} {Automatic evaluation of attribution by large language models}.
\newblock In \emph{Findings of the Association for Computational Linguistics: {{EMNLP}} 2023}, pages 4615--4635.

\bibitem[{Zha et~al.(2023)Zha, Yang, Li, and Hu}]{zha-etal-2023-alignscore}
Yuheng Zha, Yichi Yang, Ruichen Li, and Zhiting Hu. 2023.
\newblock {{AlignScore}}: {{Evaluating}} factual consistency with {{A}} unified alignment function.
\newblock In \emph{Proceedings of the 61st Annual Meeting of the Association for Computational Linguistics (Volume 1: {{Long}} Papers)}, pages 11328--11348.

\bibitem[{Zhang et~al.(2020)Zhang, Kishore, Wu, Weinberger, and Artzi}]{zhang2020bertscore}
Tianyi Zhang, Varsha Kishore, Felix Wu, Kilian~Q Weinberger, and Yoav Artzi. 2020.
\newblock \href {https://openreview.net/forum?id=SkeHuCVFDr} {{{BERTScore}}: {{Evaluating Text Generation}} with {{BERT}}}.
\newblock In \emph{8th {{International Conference}} on {{Learning Representations}}, {{ICLR}} 2020, {{Addis Ababa}}, {{Ethiopia}}, {{April}} 26-30, 2020}.

\bibitem[{Zhang et~al.(2024{\natexlab{a}})Zhang, Aliannejadi, Pei, Yuan, Huang, and Kanoulas}]{zhang2024comparative}
Weijia Zhang, Mohammad Aliannejadi, Jiahuan Pei, Yifei Yuan, Jia-Hong Huang, and Evangelos Kanoulas. 2024{\natexlab{a}}.
\newblock A comparative analysis of faithfulness metrics and humans in citation evaluation.
\newblock \emph{The First Workshop on Large Language Model for Evaluation in Information Retrieval (LLM4Eval) at SIGIR}.

\bibitem[{Zhang et~al.(2024{\natexlab{b}})Zhang, Huang, Vakulenko, Xu, Rajapakse, and Kanoulas}]{zhang2024beyond}
Weijia Zhang, Jia-Hong Huang, Svitlana Vakulenko, Yumo Xu, Thilina Rajapakse, and Evangelos Kanoulas. 2024{\natexlab{b}}.
\newblock Beyond relevant documents: A knowledge-intensive approach for query-focused summarization using large language models.
\newblock In \emph{Proceedings of the 2024 International Conference on Pattern Recognition (ICPR)}.

\bibitem[{Zhang et~al.(2024{\natexlab{c}})Zhang, Pal, Huang, Kanoulas, and de~Rijke}]{zhang2024qfmts}
Weijia Zhang, Vaishali Pal, Jia-Hong Huang, Evangelos Kanoulas, and Maarten de~Rijke. 2024{\natexlab{c}}.
\newblock \href {https://doi.org/10.48550/ARXIV.2405.05109} {{{QFMTS}}: Generating query-focused summaries over multi-table inputs}.
\newblock In \emph{Proceedings of the 27th European Conference on Artificial Intelligence (ECAI)}.

\bibitem[{Zhang et~al.(2021)Zhang, Vakulenko, Rajapakse, and Kanoulas}]{zhang2021scaling}
Weijia Zhang, Svitlana Vakulenko, Thilina Rajapakse, and Evangelos Kanoulas. 2021.
\newblock Scaling up query-focused summarization to meet open-domain question answering.
\newblock \emph{ArXiv preprint, abs/2112.07536}.

\bibitem[{Zhang et~al.(2023{\natexlab{a}})Zhang, Vakulenko, Rajapakse, Xu, and Kanoulas}]{zhang2023tackling}
Weijia Zhang, Svitlana Vakulenko, Thilina Rajapakse, Yumo Xu, and Evangelos Kanoulas. 2023{\natexlab{a}}.
\newblock Tackling query-focused summarization as a knowledge-intensive task: A pilot study.
\newblock \emph{The First Workshop on Generative Information Retrieval (Gen-IR) at SIGIR}.

\bibitem[{Zhang et~al.(2023{\natexlab{b}})Zhang, Li, Cui, Cai, Liu, Fu, Huang, Zhao, Zhang, Chen, Wang, Luu, Bi, Shi, and Shi}]{zhang2023hallucination}
Yue Zhang, Yafu Li, Leyang Cui, Deng Cai, Lemao Liu, Tingchen Fu, Xinting Huang, Enbo Zhao, Yu~Zhang, Yulong Chen, Longyue Wang, Anh~Tuan Luu, Wei Bi, Freda Shi, and Shuming Shi. 2023{\natexlab{b}}.
\newblock \href {https://doi.org/10.48550/ARXIV.2309.01219} {Siren's song in the {{AI}} ocean: {{A}} survey on hallucination in large language models}.
\newblock \emph{CoRR}, abs/2309.01219.

\bibitem[{Zhu et~al.(2024)Zhu, Huang, Rudinac, and Kanoulas}]{zhu2024enhancing}
Hongyi Zhu, Jia-Hong Huang, Stevan Rudinac, and Evangelos Kanoulas. 2024.
\newblock Enhancing interactive image retrieval with query rewriting using large language models and vision language models.
\newblock In \emph{Proceedings of the 2024 International Conference on Multimedia Retrieval}, pages 978--987.

\end{thebibliography}

\appendix

\section{Details of Prompts}
\label{app_sec:appendix_prompts}

Details of GPT-3.5 prompts used in the experiments are shown in Table \ref{tab:appendix_prompts}.

\begin{table*}[!b]
\resizebox{0.985\linewidth}{!}{
\small
\begin{tabularx}{\linewidth}{p{0.2\textwidth}<{\raggedright}l}%
\toprule[1pt]
\textbf{Prompt Name} & \textbf{Prompt Content} \\ 
\midrule
Discrete Scoring & \makecell[Xt]{\textbf{Instruction:} \\Your task is to quantify how well a provided citation supports a given statement. You should predict a \textit{discrete} score from the set $\{0, 1, 2\}$, where $0$, $1$, $2$ represent that the statement is not supported, partially supported, and fully supported, respectively. Let's think step by step. \\\\  \textbf{Statement:} \{\textit{statement}\}  \\ \textbf{Citation:} \{\textit{cited text chunk}\} \\\\ \textbf{Prediction:}} \\
\midrule
Continuous Scoring & \makecell[Xt]{\textbf{Instruction:} \\Your task is to quantify how well a provided citation supports a given statement. You should predict a \textit{continuous} score between 0 and 1 (inclusive), where 0 is not supported, 1 is fully supported, and a float value between 0 and 1 is partially supported. Let's think step by step. \\\\ \textbf{Statement:} \{\textit{statement}\} \\ \textbf{Citation:} \{\textit{cited text chunk}\} \\\\ \textbf{Prediction:}} \\
\bottomrule[1pt]
\end{tabularx}
}
\caption{Detailed prompts for discrete and continuous scoring methods.}
\label{tab:appendix_prompts}
\end{table*}

\end{document}